# Gyromagnetically Induced Transparency of Meta-Surfaces

S. Hossein Mousavi[1,2*], Alexander B. Khanikaev[3,4*‡], Jeffery Allen[5], Monica Allen[5], and Gennady Shvets[1]†

[1]*Department of Physics, University of Texas at Austin, Austin, Texas 78712, USA*
[2]*Cockrell School of Engineering, University of Texas at Austin, Austin, TX 78712 USA*
[3]*Department of Physics, Queens College of The City University of New York, Queens, New York 11367, USA and*
[4]*The Graduate Center of The City University of New York, New York, New York 10016, USA*
[5]*Sensors Directorate, Air Force Research Laboratory, Wright-Patterson AFB, OH, USA*

*These authors contributed equally to the present work*
‡khanikaev@gmail.com; †gena@physics.utexas.edu

We demonstrate that the presence of a (gyro)magnetic substrate can produce an analog of electromagnetically induced transparency in Fano-resonant meta-molecules. The simplest implementation of such gyromagnetically induced transparency (GIT) in a meta-surface, comprised of an array of resonant antenna pairs placed on a gyromagnetic substrate and illuminated by a normally incident electromagnetic wave, is analyzed. Time reversal and spatial inversion symmetry breaking introduced by the DC magnetization makes meta-molecules bi-anisotropic. This causes Fano interference between the otherwise uncoupled symmetric and anti-symmetric resonances of the meta-molecules giving rise to a sharp transmission peak through the otherwise reflective meta-surface. We show that, for an oblique wave incidence, one-way GIT can be achieved by the combination of spatial dispersion and gyromagnetic effect. These theoretically predicted phenomena pave the way to nonreciprocal switches and isolators that can be dynamically controlled by electric currents.

The concept of symmetry pervades modern physics [1, 2]. Through the conservation laws derived from various symmetries, high-level restrictions and selection rules can be derived for a variety of physical systems without any need for detailed investigations of their specific properties. Electromagnetic (EM) metamaterials [3, 4, 5] and, more specifically, Fano-resonant [6] metamaterials and meta-surfaces (FMs) [7, 8, 9, 10, 11, 12, 13, 14, 15] that are the subject of this Letter are no exception. By emulating what was originally conceived as a quantum mechanical Fano interference effect [6], FMs have been used to mimic a wide variety of other related phenomena across the electromagnetic spectrum, including slow light [16], electromagnetically induced transparency (EIT) [8, 9, 10], and electromagnetically induced absorption (EIA) [17] and polarization conversion [18]. Spectrally sharp EM resonances and their associated strong EM field enhancements in FMs are beneficial for a broad range of applications, such as low-threshold lasing [19], enhanced nonlinear response [20, 21, 22], dynamic tuning [23, 24, 25, 26, 27, 28, 25], and sensing and biosensing [29, 30, 31, 32, 33]. The spatial symmetries of electric charge distribution on the metamaterial's surface determine whether the EM resonance is "bright" (radiatively coupled to) or "dark" (radiatively de-coupled from) the EM continuum. As we demonstrate in this letter, other symmetries and their breaking can also be crucial to determine the properties of EM resonances and enable their mutual coupling, which in turn can give rise to EM Fano interferences.

We consider a meta-surface formed by a two-dimensional array of double-antenna meta-molecules [34] [Fig.1(a)]. Such meta-molecules support two low-frequency resonances formed by hybridization of the dipolar modes of the individual antennas that have distinct symmetric and anti-symmetric charge/current distributions responsible for their strongly disparate radiative coupling. The scattering of light by the symmetric mode is dominated by the electric dipolar moment of the meta-molecules ($d_y$) along the $y$-direction, leading to its strong radiative coupling and its "bright" character. The anti-symmetric mode, in contrast, is dominated by the magnetic dipolar ($m_z$) and electric quadrupolar ($q_{xy}$) moments, leading to its "darkness", i.e. its complete decoupling from the normally incident EM wave [see Fig.1(b)]. Fano interference arises from the coupling between the symmetric and anti-symmetric resonances, which are a consequence of coupling between: (*i*) the electric and magnetic dipolar moments $d_y \sim \kappa_{dm} m_z$, *i.e.* magneto-electric coupling [ 35], and (*ii*) the electric dipolar and electric quadrupolar moments $d_y \sim \kappa_{dq} q_{xy}$, essentially the spatial dispersion. Such coupling of multipolar moments makes the meta-molecules bianisotropic and occurs when the spatial symmetry of the meta-molecules is reduced [ 36]. Two well-known mechanisms to induce bianisotropy is by breaking of spatial symmetry: *i) intrinsic* [ 9, 7, 17], when the meta-molecule's geometry is distorted, e.g., by making the antennas unequal [ 7, 16, 33]), and, *ii) extrinsic*, by utilizing obliquely incident waves with finite component of wavenumber in the *x*-direction [ 37, 38, 39, 40]). In this letter, we propose a third symmetry-breaking mechanism of inducing magneto-electric coupling. By adding a magnetized gyromagnetic (GM) ferrite substrate to the structure (without either intrinsic or extrinsic mechanism) that removes both the time-reversal (TR) and spatial-inversion (SI) symmetries, we demonstrate the phenomenon of *gyromagnetically induced Fano resonance.* This phenomenon is then utilized for designing novel tunable metamaterials exhibiting *nonreciprocal (and one-way) Fano resonances* and *Gyromagnetically Induced Transparency* (GIT). Earlier works in metamaterials/photonics utilized the natural magnetic response of ferrites to achieve and control negative refractive index [ 41, 42, 43] and nonreciprocity [ 44, 45, 46, 47, 48, 49]. The focus of this letter is on demonstrating how the gyromagnetic response of materials induces Fano interferences.

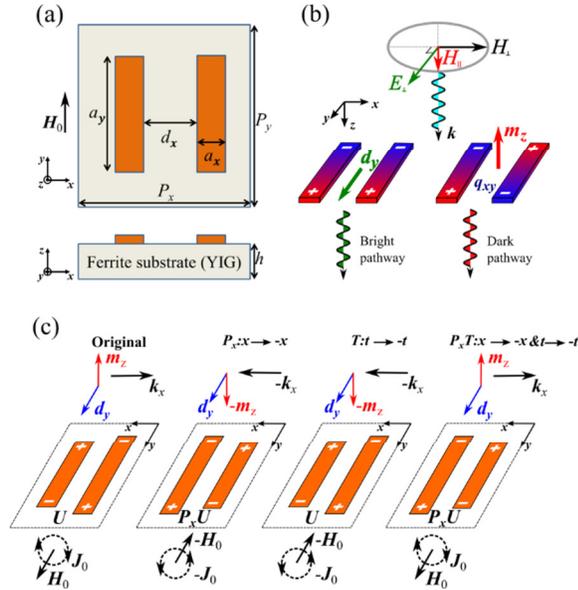

**Fig. 1 (Color online).** (a) Fano-resonant double-antenna array on top of a ferrite substrate. Dimensions are $P_x = P_y = P = 3.5$ cm, $d_x = 3.5$ mm, $a_y = 2.45$ cm, $a_x = 3.5$ mm, $h = 8.75$ mm. (b) Charge distribution and excitation of symmetric and anti-symmetric modes of the meta-molecule in a magnetized medium. (c) Symmetry operations explaining different mechanisms of magneto-electric coupling.

The EM response of a bi-resonant FM to the incident EM wave with the amplitude $E^{in}$ shown in Fig.1 is described within the framework of temporal coupled-mode theory (CMT)[ 50, 51, 52]:

$$\left[\frac{d}{dt} - i\begin{pmatrix} \widetilde{\omega}_d & -i\kappa^* \\ i\kappa & \widetilde{\omega}_m \end{pmatrix}\right]\begin{pmatrix} d_y \\ m_z \end{pmatrix} = \begin{pmatrix} i\alpha_E \\ 0 \end{pmatrix} S^{in}, \qquad (1)$$

Note that the anti-symmetric mode can be fully characterized by one of its multipolar amplitudes, $q_{xy}$ or $m_z$, which are proportional to each other. To avoid redundancy, we will describe the dark anti-symmetric mode in terms of the magnetic dipolar moment $m_z$. Thus, in Eq.(1), $d_y$ and $m_z$ represent the complex-valued spectral amplitudes of the symmetric and anti-symmetric modes, respectively, $\kappa$ is the coupling coefficient between these modes, and $S^{in}$ characterizes the electric field amplitude of an incident s-polarized ($\boldsymbol{E^{in}}||\boldsymbol{y}$) wave. The case of Voigt geometry ($\boldsymbol{H_0}||\boldsymbol{y}$) is considered here. Note that the presence of the imaginary unit in front of the coupling coefficient in Eq.(1) is introduced to consider for the different behavior of the electric and magnetic moments under TR [$T(d_y) = d_y$, $T(m_z) = -m_z$]. The time dependence $e^{i\omega t}$ is assumed for $d_y, m_z$ and $S^{in}$. The amplitudes $d_y$, $m_z$ and $S^{in} = \sqrt{\frac{\epsilon_0 c}{2}} E^{in} e^{i\omega t}$ are normalized so that $|d_y|^2$ and $|m_z|^2$ are equal to the energy per unit area stored by the corresponding modes, and $|S^{in}|^2$ is equal to the irradiance [51, 52] of the incident field. The frequencies (lifetimes) of the two modes are given by $\omega_{d,m}$ ($\tau_{d,m}$), and $\widetilde{\omega}_{d,m} \equiv \omega_{d,m} + i\tau_{d,m}^{-1}$. In the absence of the non-radiative loss assumed here, $\tau_m^{-1} = 0, \tau_d^{-1} = \alpha_E^2$ [50], where $\alpha_E$ is the radiative coupling parameter. The complex-valued reflection coefficient $r = -\alpha_E^* d_y/S^{in}$ [50] found by solving Eq.(1) is given by

$$r = -\frac{|\alpha_E|^2}{(\omega-\widetilde{\omega}_d)-\frac{|\kappa|^2}{(\omega-\widetilde{\omega}_m)}}, \qquad (2)$$

and the transmission coefficient is calculated as $t = 1 + r$ [50, 38] assuming an infinitely thin electric metasurface [53].

It follows from Eq.(2) that the Fano resonance emerges due to the coupling between the symmetric and anti-symmetric modes when the coupling coefficient is finite. In general, $\kappa$ is a function of the unit cell structure ($U$), magnetic field ($H_0$), and the wavenumber ($k_x$): $\kappa \equiv \kappa(H_0, U, k_x)$. By applying simple symmetry arguments to the CMT Eq.(1), we show that $\kappa \neq 0$ when the GM substrate is magnetized, even for a symmetric unit cell and at normal incidence.

Three symmetry operations are applied to the meta-molecule and underlying substrate: SI ($P_x: x \to -x$), TR ($T: t \to -t$), and their combination $P_x T$ ($x \to -x, t \to -t$). DC magnetic field transforms under these symmetry operations according to $T(H_0) = -H_0$, $P_x(H_0) = -H_0$, and $P_x T(H_0) = H_0$, as schematically explained in Fig.1(c). The same holds for the wave-number $k_x$, and $P_x(U) = U$ holds for a symmetric unit cell. Applying these symmetry transformations to the symmetric and anti-symmetric modes, as shown in Fig.1(c), we are looking for the appropriate form of the coupling coefficient between the dipolar moment of the bright symmetric mode with the magnetic moment $\omega m_z \sim i\kappa d_y$ of the dark anti-symmetric mode. Assuming that the contribution of the three mechanisms of coupling: (i) intrinsic due to reduced symmetry of the meta-molecule [$P_x(U) \neq U$], (ii) extrinsic due to oblique incidence ($k_x \neq 0$), and (iii) gyromagnetic due to the presence of magnetization ($H_0 \neq 0$), are small, independent and therefore additive, we are looking for the coupling coefficients of the form $\kappa(U, k_x, H_0) = \widetilde{F} + \widetilde{A}k_x + \widetilde{G}H_0$, where the three complex-valued coefficients $\widetilde{F}$, $\widetilde{A}$, and $\widetilde{G}$ represent the three corresponding contributions to the coupling coefficient. Here $\widetilde{F}$ represents the degree of asymmetry of a unit cell ($P_x(U) \neq U$), $\widetilde{A}$ characterizes the strength of spatial dispersion [54, 39, 40], and $\widetilde{G}$ quantifies the gyromagnetic response.

It is clear from comparison of the original and transformed systems in Fig.1(c) that (i) $P_x$- and $T$-transformed systems are physically identical to the original one when TR symmetry is present ($H_0 = 0$), and (ii) $P_x T$-transformed system is identical to the original one when the unit cell is symmetric, and therefore are described by the same Eq.(1). These two observations immediately lead to the following constraints on $\kappa$ (refer to Supplementary section) [39, 55]:

$$\kappa(H_0 = 0, P_x(U), -k_x) = -\kappa(H_0 = 0, U, k_x), \quad (3a)$$
$$\kappa^*(H_0 = 0, U, -k_x) = \kappa(H_0 = 0, U, k_x), \quad (3b)$$
$$\kappa^*(H_0, P_x(U) = U, k_x) = -\kappa(H_0, P_x(U) = U, k_x). \quad (3c)$$

Here the complex conjugation arises from the TR operation [55]. Equation 3(a) implies that $\kappa$ is an odd function of $k_x$ and the unit cell asymmetry. Equation 3(b) represents the condition of reciprocity in the presence of TR symmetry, while Eq.3(c) describes the effect of TR and SI symmetry removal due to the DC magnetic field [56]. From the constraints Eq.(3), one can easily verify that the following expression for the $\kappa\,(H_0, U, k_x)$ satisfies all the three conditions:

$$\kappa(H_0, U, k_x) = F + i(Ak_x + GH_0) \quad (4)$$

where the three coefficients $F$, $A$, and $G$ have reduced to the real-valued numbers, and, as before, represent the intrinsic, extrinsic, and gyromagnetic contributions to bianisotropy, respectively.

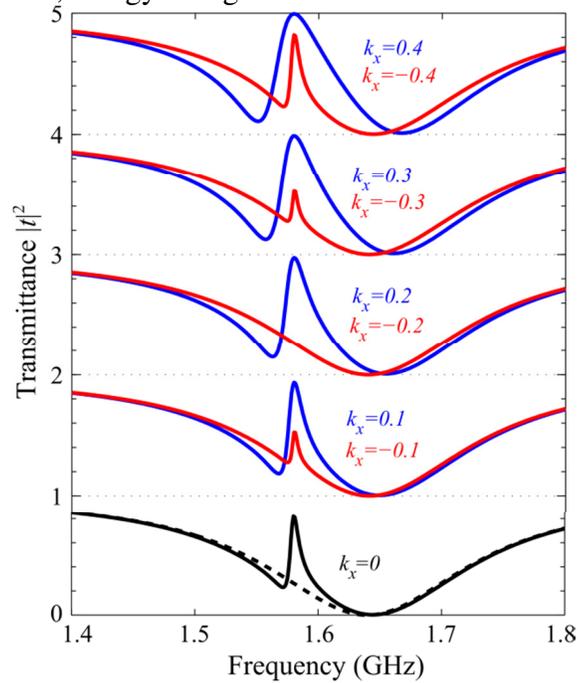

**Fig. 2 (Color online).** Transmission spectra illustrating GIT and one-way Fano resonance as predicted by Eq.(2) for different incidence angles ($k_x$) for the case of symmetric ($F = 0$) meta-molecules. Black dashed and solid lines show nonmagnetic ($H_0 = 0$) and magnetic ($H_0 \neq 0$) substrates, respectively. Parameters used: $|\alpha_E|^2 = 0.1$ GHz, $\omega_s = 1.64$ GHz, $\tau_s^{-1} = 0.1$ GHz, $\omega_a = 1.58$ GHz, $GH_0 = 0.016$ GHz, $A=0.08$ GHz · m.

Several important conclusions can be drawn from Eq.(4): *(i)* if TR symmetry is not broken by the imposed magnetic field, the magneto-electric response can be decomposed into intrinsic and extrinsic parts [39]: $\kappa_{int} + \kappa_{ext}(k_x) = F + iAk_x$; *(ii)* breaking of the TR and SI by the substrate magnetization is by itself sufficient for inducing the bianisotropy of a symmetric, normally illuminated meta-molecule: $\kappa_{gyr} = iGH_0$, thereby causing Fano interference and GIT; and *(iii)* the extrinsic and gyromagnetic contributions exactly cancel each other in the case of a symmetric meta-molecule if $k_x = -GH_0/A$. Property *(iii)* implies that there exists one specific incidence angle $\theta_c = -\sin^{-1}[cGH_0/\omega A]$ for which the Fano interference is identically cancelled in *forward* (but not *backward*) direction giving rise to *one-way Fano interference*, and that takes place for $\kappa(k_x) = i(GH_0 + Ak_x) = 0$ [but $\kappa(-k_x) = i(GH_0 - Ak_x) \neq 0$]. Note that such cancellation is mathematically possible only for $F = 0$ (*i.e.,* symmetric meta-molecules).

While the first property is well-known, the last two predictions constitute the main result of this letter and are illustrated in Fig.2 by the use of the CMT Eqs. (2) and (4). Black dashed and solid lines in Fig.2 correspond to the cases of nonmagnetic ($H_0 = 0$) and magnetic ($H_0 \neq 0$) substrates and illustrate the property *(ii),* predicting the appearance of a sharp spectral feature, indicative of the Fano resonance and GIT at normal incidence ($k_x = 0$) when the substrate is magnetized. For the cases of oblique incidence ($k_x \neq 0$), Fig.2 also demonstrates strongly nonreciprocal electromagnetic response of the meta-surface $|t(k_x)|^2 \neq |t(-k_x)|^2$. A particular case illustrating property *(iii)* of a one-way GIT corresponds to the situation when the gyromagnetic and extrinsic contributions to the bianisotropy exactly cancel each other in forward direction ($k_x = 0.2$ m$^{-1}$) but not for the backward direction ($k_x = -0.2$ m$^{-1}$), where these contributions add up to augment the Fano feature.

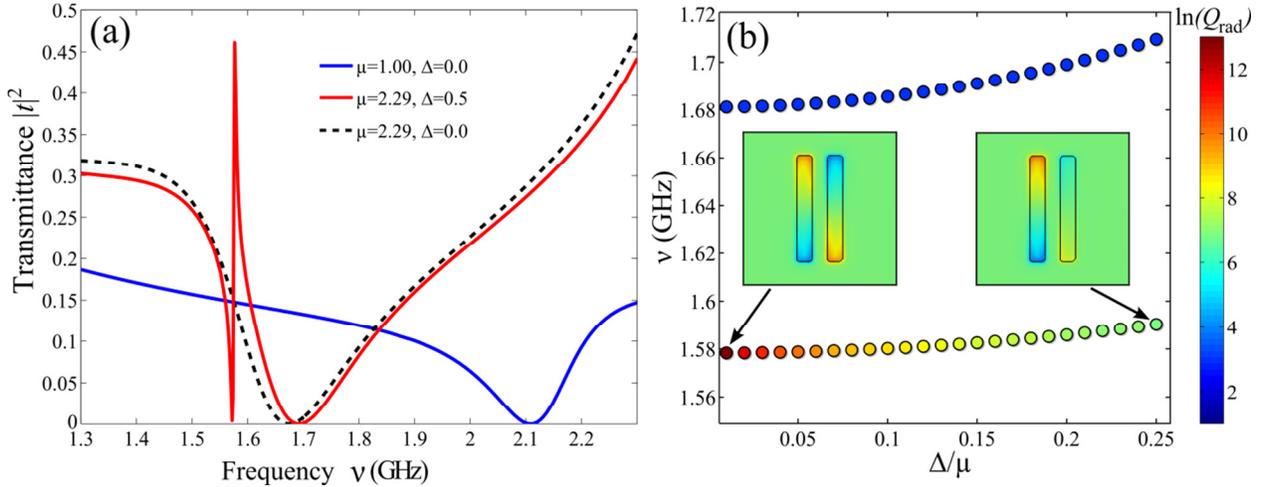

**Fig. 3 (Color online).** GIT at normally incidence. (a) Transmission spectra without (blue) and with (red) substrate magnetization. The narrow GIT peak is due to the off-diagonal component of the substrate's magnetic permeability tensor no peak for $\Delta = 0$ (dashed line). (b) Spectral position and radiative quality factor $Q_{rad}$ (colorbar) of the symmetric (high-frequency) and anti-symmetric (low-frequency) modes as a function of gyromagnetic activity $\Delta/\mu$. Insets: charge distribution for two values of $\Delta/\mu$. Geometric parameters: same as Fig.1. Substrate parameters: $\epsilon = 15$, $\mu = 1 + \frac{\omega_m(\omega_0 + i\alpha)}{(\omega_0 + i\alpha)^2 - \omega^2}$, $\Delta = \frac{\omega_m \omega}{(\omega_0 + i\alpha)^2 - \omega^2}$, $\omega_0 = \gamma H_{0i}$, $\omega_m = 4\pi\gamma M_s$, and $\alpha = \gamma \Delta H/2$. For the $Y_3Fe_5O_{12}$ ferrite medium [57]: $4\pi M_s = 1750$ [G], $\gamma = 1.760$e11 $\left[\frac{\text{rad}}{\text{sT}}\right]$, and $H_{0i} = 1600$ [Oe]. The ferrite is assumed to be lossless ($\alpha = 0$ and $\Delta H = 0$) in these simulations.

The above implications, GIT and one-way Fano interference, are confirmed with first-principle numerical simulations (COMSOL Multiphysics) of the FM on top of YIG substrate. The GIT phenomenon is illustrated by Fig.3(a), where normal-incidence transmission spectra are plotted without (blue line) and with (red line) magnetization in the substrate. The components of the permittivity tensor $\hat{\mu} = \mu(\vec{e}_x\vec{e}_x + \vec{e}_z\vec{e}_z) + \vec{e}_y\vec{e}_y + i\Delta(\vec{e}_x\vec{e}_z - \vec{e}_z\vec{e}_x)$ of YIG-based substrate are given in Fig.3 caption. Without the magnetic field, the spectrum exhibits a single broad transmission dip at $\nu_d = 2.1$ GHz corresponding to the resonant excitation of the bright symmetric mode. A dramatically different Fano-like transmission spectrum is found for finite DC magnetic field in the substrate. A sharp asymmetric transmission peak emerges at the frequency of the dark anti-symmetric mode $\nu_m \equiv \frac{\omega_m}{2\pi} \approx 1.59$ GHz, which is superimposed on a broad transmission dip centered at the bright mode frequency $v_d = 1.7$ GHz (modified due to the magnetization). Since the transmission peak occurs without extrinsic or intrinsic reduction of the spatial symmetry, but, in the presence of the magnetized ferrite, we conclude that the transparency is caused by breaking TR and SI symmetries by the DC magnetic field. The overall spectral redshift and larger transmission for the magnetized substrate is due to the higher magnetic permeability $\mu_{zz} \equiv \mu = 2.29$ and better impedance matching between the substrate and vacuum.

We also confirm that the observed GIT is due to the off-diagonal $\Delta$ component of the substrate's permeability tensor $\hat{\mu}$ by performing test calculations that artificially set $\Delta = 0$ while keeping $\mu = 2.29$. The simulation results, shown in Fig.3a by the dashed line, show a transmission spectrum that does not exhibit Fano resonance. Remarkably, the bandwidth of the GIT peak can be continuously tuned by increasing magnetization of the substrate. Fig.3b shows how such controllability of the anti-symmetric mode's quality factor $Q$ can be achieved by varying gyromagnetic activity $\Delta/\mu$. In this figure, the color-coded plots of the complex eigenvalue frequencies $\widetilde{\omega}_{d,m} \equiv 2\pi v_{d,m}(1 + iQ_{d,m}^{-1})$ of the symmetric/anti-symmetric modes, obtained using COMSOL eigenvalue simulations, are presented. We observe that a small variation of the parameter $\Delta/\mu$ (from $\Delta/\mu = 0.1$ to $\Delta/\mu = 0.25$) results in a dramatic change of the radiative quality factor (from $Q_m = 9357$ to $Q_m = 587$) of the anti-symmetric mode. This change is associated with increased radiative coupling of the anti-symmetric mode caused by the modification of its field profile. As can be seen from the Fig.3(b) inset, as $\Delta/\mu$ increases, the dark mode loses its perfectly anti-symmetric profile and acquires a significant dipolar component. In contrast to this [and in consistence with Fig.3(a)], the quality factor of the bright symmetric mode is almost independent of $\Delta/\mu$ and remains much lower than that of the dark mode.

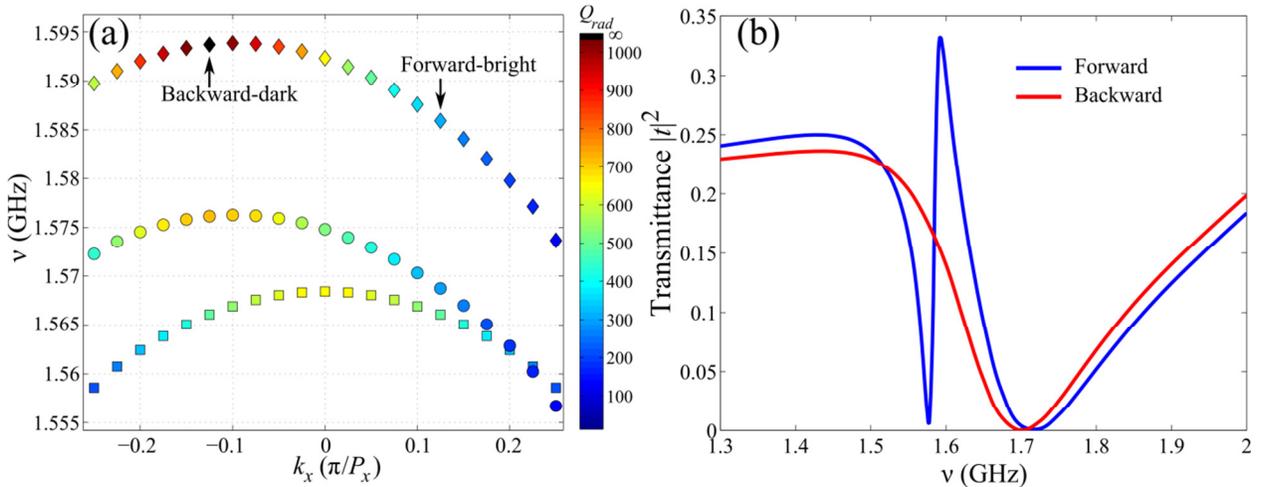

**Fig. 4 (Color online).** (a) Complex photonic band of the anti-symmetric mode. Diamonds: symmetric antennas on a gyromagnetic ($\Delta/\mu = 0.2$) substrate. Squares: asymmetric antennas (the right antenna is 2.5% longer) on a nonmagnetic ($\Delta/\mu = 0$) substrate. Circles: asymmetric antennas on a gyromagnetic substrate. (b) Transmission spectrum illustrating one-way Fano-resonance at the incidence angle $\theta = 24^{\circ}$. The structure parameters are as before, except $\Delta H=30$ [Oe] is assumed in (b) to account for ferrite losses.

We illustrate one-way (nonreciprocal) Fano interference and GIT, by simulating transmission of obliquely incident EM waves through the FM. To determine the critical angle $\theta_c$, we carried out eigenvalue simulations using COMSOL to determine complex band diagrams $\widetilde{\omega}_{d,m}(k_x)$ for a fixed value of $\Delta/\mu = 0.2$. The color-coded band structures in Fig.4(a) show the anti-symmetric mode's radiative quality factor $Q_m(k_x)$ for three combinations of substrates and meta-molecule geometries: (i) symmetric (equal length) antennas and a gyromagnetic substrate (diamonds); (ii) asymmetric (different length) antennas and a non-gyromagnetic substrate (squares), and (iii) asymmetric antennas and a gyromagnetic substrate (circles). As expected for cases (i) and (iii), the anti-symmetric mode exhibits nonreciprocity of its resonant frequency, $\nu_m(k_x) \neq \nu_m(-k_x)$, and its quality factor $Q_m(k_x) \neq Q_m(-k_x)$. Moreover, in the case (i), the one-way cancelation of the Fano interference is observed. Specifically, for $|k_x| \approx 0.12\pi/P$ (corresponding to the critical angle $\theta_c \approx 24°$) the anti-symmetric mode is completely decoupled (*i.e.,* the radiative $Q_m(k_x) \to \infty$ diverges for $k_x = -0.12\pi/P$) from the backward-incident radiation, but is coupled to the forward-incident radiation ($k_x = 0.12\pi/P$). Note that such cancelation does not occur in the case (iii) corresponding to the asymmetric meta-molecules: the radiative quality factor remains finite for all values of $k_x$.

COMSOL calculations of the transmission spectra for realistic losses indeed confirm the one-way GIT at the oblique incidence angle $\theta_c = 24°$ obtained from the above eigenvalue simulations. The transmission spectra plotted in Fig.4(b) show a distinct narrow-band transmission peak in the forward direction (blue line) but not in the backward direction (red line).

In conclusion, we have used simple symmetry considerations to predict and numerically demonstrate two phenomena that occur in meta-surfaces when symmetry of the system is reduced by a gyromagnetic substrate: *gyromagnetically induced transparency* and *nonreciprocal Fano interference*. These phenomena hold significant promise for practical applications such as the dynamic control of resonant EM interactions using magnetic fields produced by the external currents, mitigation of co-site interference and improving isolation. Spectral positions, radiative lifetimes and quality factors of Fano resonances can be controlled by the magnitude of the external magnetic field. While similar tunability may be achieved with other methods, the approach based on gyromagnetically induced coupling to dark sub-radiant resonances proposed in this letter is unique because of its nonreciprocal nature. This class of effects may lead to a new generation of tunable and nonreciprocal Fano resonant systems for various applications where strong field enhancement, tunability, and nonreciprocity are simultaneously required. One-way absorbers, one-way sensors, and one-way cloaking elements are just a few examples of such applications.

**Acknowledgment**
The authors (JWA and MSA) are thankful for the funding support through AFOSR Lab Task 13RY02COR (PO: Dr. Harold Weinstock). SHM, ABK and GS acknowledge support from the Air Force Research Laboratory (AFRL/RY), through the Advanced Materials, Manufacturing and Testing Information Analysis Center (AMMTIAC) contract with Alion Science and Technology Contract # FA4600-06-D003.